\documentclass{article}
\usepackage[english]{babel} 

\usepackage{graphicx}

\usepackage{color}

\begin{document}

\newcommand{\rum}{\rule{0.5pt}{0pt}}
\newcommand{\rub}{\rule{1pt}{0pt}}
\newcommand{\rim}{\rule{0.3pt}{0pt}}
\newcommand{\numtimes}{\mbox{\raisebox{1.5pt}{${\scriptscriptstyle \rum\times}$}}}
\newcommand{\numtimess}{\mbox{\raisebox{1.0pt}{${\scriptscriptstyle \rum\times}$}}}
\newcommand{\Boldsq}{\vbox{\hrule height 0.7pt
\hbox{\vrule width 0.7pt \phantom{\footnotesize T}%
\vrule width 0.7pt}\hrule height 0.7pt}}
\newcommand{\two}{$\raise.5ex\hbox{$\scriptstyle 1$}\kern-.1em/
\kern-.15em\lower.25ex\hbox{$\scriptstyle 2$}$}

\renewcommand{\refname}{References}
\renewcommand{\tablename}{\small Table}
\renewcommand{\figurename}{\small Fig.}
\renewcommand{\contentsname}{Contents}

\begin{center}
{\Large\bf Reply to ``Isotropy of Speed of Light" by Casta\~{n}o and Hawkins, \\ arXiv:1103.1620\rule{0pt}{13pt}}\par

\bigskip
Reginald T. Cahill \\ 
{\small\it  School of Chemical and  Physical Sciences, Flinders University,
Adelaide 5001, Australia\rule{0pt}{15pt}}\\
\raisebox{+1pt}{\footnotesize E-mail: Reg.Cahill@flinders.edu.au}\par
\bigskip

{\small\parbox{11cm}{%
In ``Isotropy of Speed of Light" by Casta\~{n}o and Hawkins, arXiv:1103.1620, it is claimed, using a flawed theoretical argument, that the speed of light must necessarily be isotropic, independent even of experiment. The key false assumption made is that the round trip time must always be invariant wrt change of direction of the light path. This is shown to be  false. More importantly the anisotropy of the speed of light has been repeatedly detected in experiments, beginning with Michelson and Morley in 1887, and with the most recent data being from spacecraft earth-flyby Doppler shift data.  Similar misunderstandings critically  affect the designs for LIGO and LISA.\rule[0pt]{0pt}{0pt}}}\medskip
\end{center}

\setcounter{section}{0}
\setcounter{equation}{0}
\setcounter{figure}{0}
\setcounter{table}{0}
\setcounter{page}{1}


\section{Introduction}
In ``Isotropy of Speed of Light" by Casta\~{n}o and Hawkins, arXiv:1103.1620 \cite{CH}, it is claimed, using a flawed theoretical argument, that the speed of light must necessarily be isotropic, independent even of experiment. The key false assumption made is that the round trip time must always be invariant wrt change of direction of the light path, although no reasons for that assumption are given.  This is shown to be  false. More importantly the anisotropy of the speed of light has been repeatedly detected in experiments, beginning with Michelson and Morley in 1887 \cite{MM,MMCK,MMC},  and with the most recent data being from spacecraft earth-flyby Doppler shift data
 \cite{CahillNASA, And2008}. Other experiments were reported in \cite{Miller,Illingworth,Joos,Jaseja,Torr,Krisher,DeWitte, Munera}.
 It is also noted that the misunderstanding of the physics associated with the anisotropy of the speed of light explains why LIGO and related instruments have failed to detect gravitational waves, while the planned LISA  space-based  detector will be excessively sensitive.

 \begin{figure}[ht]
\vspace{15mm}\setlength{\unitlength}{1.0mm}
\hspace{47mm}\begin{picture}(20,30)\put(-5,0){\line(1,0){65}}
\put(0,-3){{\bf A}}
\put(6,20){{\bf L}}
\put(26,20){{\bf L}}
\put(46,20){{\bf L}}
\put(20,41){{\bf A}$'$}

\color{green}
\put(31.8,0){\line(1,5){8.0}}
\put(32,0){\line(1,5){8.0}}
\put(32.2,0){\line(1,5){8.0}}

\put(0,0){\line(1,1){39.6}}
\put(0.2,0){\line(1,1){39.6}}
\put(-0.2,0){\line(1,1){39.6}}

\color{black}
\put(30,-3){{\bf C}}

\put(19,-3){{\bf B}}
\put(23,+1.5){{$\theta$}}

\put(35,42){\line(2,-1){9}}
\put(39,41){{\bf B}$'$}

\put(44,8){\vector(1,0){10}}

\put(20,0){\line(1,2){20}}

\put(0.0,0){\line(1,2){19.8}}
\put(15,42){\line(2,-1){9}}

\put(31.6,0){\line(1,2){19.8}}
\put(47,42){\line(2,-1){9}}
\put(52,41){{\bf C}$'$}

\put(48,10){{${\bf  v}$}}

\put(14.7,15){\vector(1,1){2}}
\put(34.9,15){\vector(-1,-4){0.5}}

\end{picture}

\vspace{3mm}
\caption{\small{Photon bounce trajectory in reference frame of 3-space, so speed of light is $c$ in this frame.     The source is at successive locations A, B, C,  at times $t_A, t_B, t_C$, and the retroreflector is at corresponding locations A$'$, B$'$, C$'$  at the same respective times $t_A, t_B, t_C.$ Source-retroreflector  separation distance is $L$,  and has angle $\theta$ wrt   velocity ${\bf v}$ of source and reflector, and shown at three successive times: (i) when photon pulse  leaves  $A$ (ii) when photon pulse is reflected at retroreflector at $B'$, and (iii) when photon pulse  returns to source at $C$.  The round trip time depends on whether the retroreflector is support by a rod, of rest-length $L$, or not, as the rod is subject to Fitzgerald-Lorentz contraction.  The round trip time is only  independent  of $\theta$    in the case of  a rod supporting the source-retroreflector separation, and then only if the light propagates through a vacuum.}
\label{fig:Bounce}}
\end{figure}
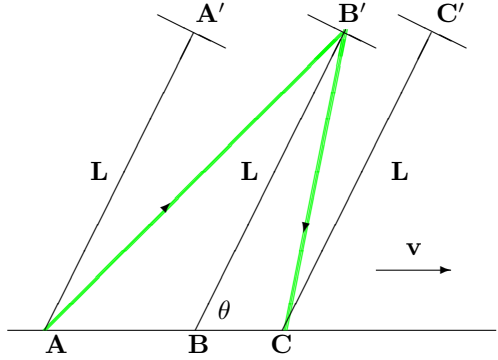

\section{Light Travel Times}

 Fig.\ref{fig:Bounce}  shows  photon bounce   trajectory in reference frame fixed in space, and so the light has speed $c$, and with source and retroreflector in motion through space with velocity ${\bf v}$.  
 Define $t_{AB}=t_{B}-t_A$ and  $t_{BC}=t_{C}-t_{B}$. The distance $AB$ is $vt_{AB}$ and distance   $BC$ is $v t_{BC}$. The total  photon travel time is  $t_{AC}=t_ {AB}+t_{BC}$.  In the case of  there being no rod supporting the source and retroreflector, so that they are merely co-moving through space, then the distance  between source and retroreflector is $L$. 
 Applying the cosine  theorem to triangles $ABB'$ and $CBB'$  we obtain
\begin{equation}\label{eqn:EM}
t_{AB}=\frac{v L\cos(\theta)+\sqrt{v^2L^2\cos^2(\theta)+L^2(c^2-v^2)}}{(c^2-v^2)}
\end{equation}
\begin{equation}\label{eqn:ME}
t_{BC}=\frac{-v L\cos(\theta)+\sqrt{v^2L^2\cos^2(\theta)+L^2(c^2-v^2)}}{(c^2-v^2)}
\end{equation}
Then to $O(v^2/c^2)$ 
\begin{equation}\label{eqn:EME}
t_{AC}=\frac{2L}{c}+\frac{L v^2(1+ \cos^2(\theta))}{c^3}+.....
\end{equation}
Hence we see that the round-trip travel time depends on orientation, contrary to the false assumption in \cite{CH}.
However if there was a solid rod separating source and reflector,  as in one arm of a  vacuum-mode Michelson interferometer, then there would be a Lorentz contraction of that rod, and in the above we need to make the replacement  $L\rightarrow L\sqrt{1-v^2 \cos^2(\theta)/c^2}$, giving $t_{AC}=2L/c+Lv^2/c^3$ to $O(v^2/c^2)$.   And  then there is no dependence of the travel time on orientation.

If, as well as a rod separating the source and retroreflector, a gas is present, and if we use the approximation  $c\rightarrow c/n$, with $n$ the refractive index, but only in the light propagation terms, but not in the Lorentz contraction, we obtain
$$t_{AC}=\frac{2Ln}{c}+\frac{Ln(n^2+(n^2-1)\cos(\theta)^2)v^2}{c^3 }\mbox{\   to  \ } O(v^2/c^2)$$
and then an angle dependence is restored, but only when $n\neq 1$. 
Further if we have two orthogonal arms of a gas-mode Michelson interferometer, the travel time in the 2nd arm is, using $\theta\rightarrow \theta+\pi/2$ for this arm,
$$t_{\perp AC}=\frac{2Ln}{c}+\frac{Ln(n^2+(n^2-1)\sin(\theta)^2)v^2}{c^3} \mbox{\   to  \ } O(v^2/c^2)$$ 
and then the difference in travel times, as measured by fringe shifts, is 
\begin{equation}\Delta t=t_{AC}-t_{\perp AC}=\frac{Ln(n^2-1)\cos(2\theta)v^2}{c^3} \mbox{\   to  \ } O(v^2/c^2)\label{eqn:dt}\end{equation}
However the above analysis does not correspond to how the interferometer is actually operated. That analysis does not actually predict fringe shifts, for the  field of view would be uniformly illuminated, and the observed effect would be a changing level of luminosity rather than fringe shifts. As Michelson and Miller knew, the mirrors must be made slightly non-orthogonal with the degree of non-orthogonality determining how many fringe shifts were visible in the field of view. Miller exper\-i\-mented with this effect to determine a comfortable number of fringes: not too few and not too many.  Hicks  deve\-lop\-ed a theory for this effect -- however it is not necessary to be aware of the details of this analysis in using the interferometer: the non-orthogonality   reduces the symmetry of the device, and instead of having period of 180$^\circ$ the symmetry now has a period of 360$^\circ$, so that to (\ref{eqn:dt})  we must add the extra term $a\cos(\theta-\beta)$ in
 \begin{equation}
\Delta t=k^2\frac{L(1+e\theta)v_P^2}{c^3}\cos\bigl(2(\theta-\psi)\bigr)+a(1+e\theta)\cos(\theta-\beta)+f
\label{eqn:dtHick}\end{equation} 
where $k^2=n(n^2-1)$.  
\begin{figure}[t]
\vspace{-0mm}
\hspace{35mm}\includegraphics[scale=0.8]{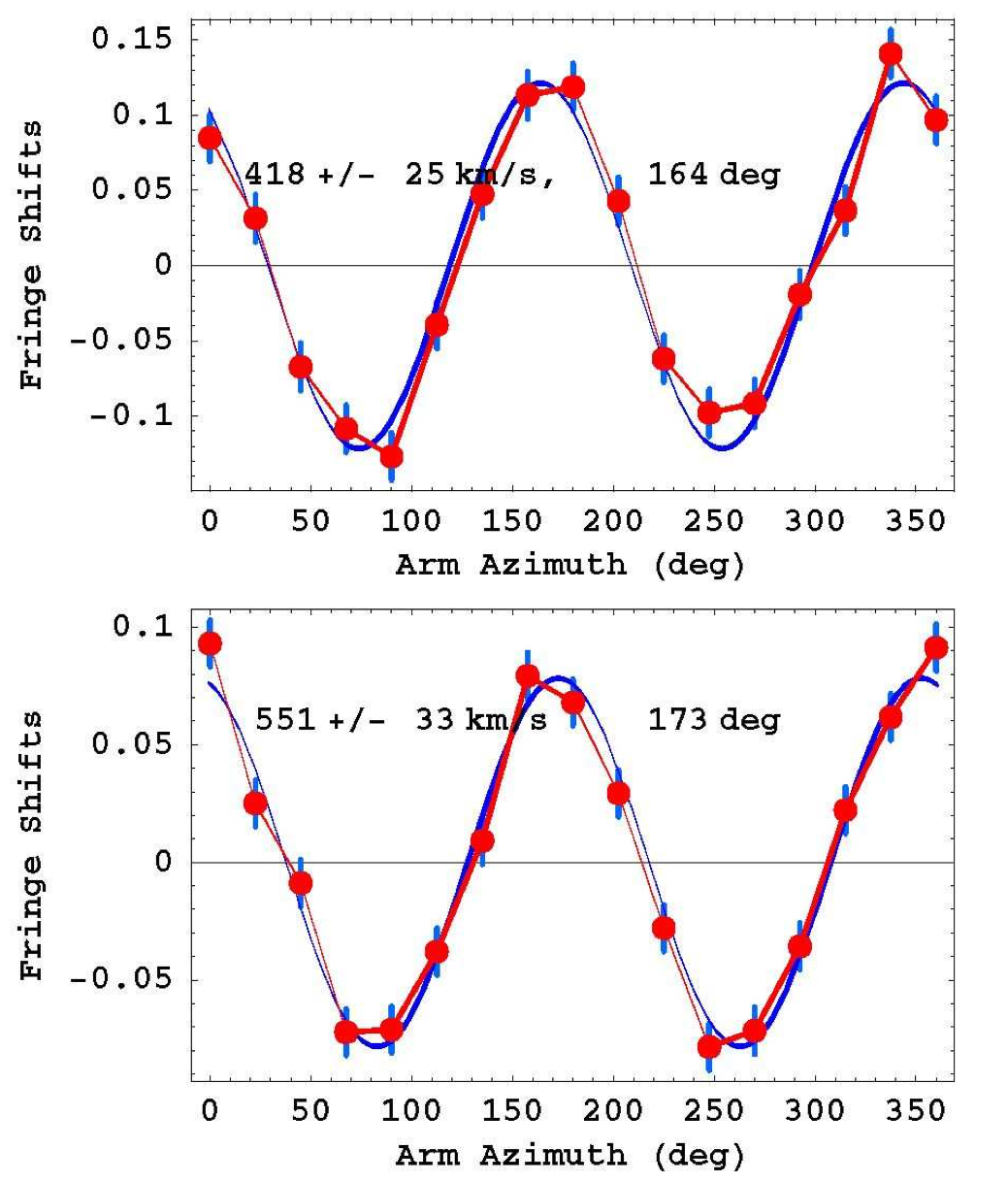}
\caption{\small {(a) A typical  Miller  averaged-data from September 16, 1925, $4^h 40^\prime$  Local Sidereal Time (LST) - an average of data from 20 turns of the gas-mode (air) Michelson interferometer. Plot and data after fitting and then subtracting both the temperature drift and Hicks effects from both, leaving the expected sinusoidal form.  The error bars are determined as the rms error in this fitting procedure, and show how exceptionally small were the errors, and which agree with Miller's claim for the errors. (b) Best result  from the Michelson-Morley 1887 data - an average of 6 turns, at  $7^h$  LST on July  11, 1887.  Again the rms error is remarkably small.  In both cases the indicated speed is  $v_P$ - the 3-space speed projected onto the plane of the interferometer. The angle is  the azimuth of the  3-space speed projection at the particular LST.  The speed fluctuations from day to day significantly exceed these errors, and reveal the existence of 3-space flow turbulence - i.e gravitational waves.}}
\label{fig:MillerMMPlots}\end{figure}
The term $1+e\theta$ models the temperature effects, namely that as the arms are uniformly rotated, one rotation taking several minutes, there will be a temperature induced change in the length of the arms. If the temperature effects are linear in time, as they would be for short time intervals, then they are linear in $\theta$. In the  Hick's term the parameter $a$ is proportional to the length of the arms, and so also has the temperature factor. The term $f$ simply models any offset  effect.
Michelson-Morley and Miller took these two effects  into account  when analysing their data. Fig.\ref{fig:MillerMMPlots} shows just such fringe shifts, clearly demonstrating the  $\cos(2\theta)$ signature, in both the Michelson-Morley and Miller gas-mode interferometer experiments.
 
Applying the above to a  laboratory  vacuum-mode    Michelson interferometer with $n=1$, as in \cite{cavities},  implies that it is unable to detect light-speed anisotropy.  This is a basic   design flaw in laboratory vacuum-mode Michelson interferometers.  The ``null" results from such devices are usually incorrectly reported as proof of the invariance of the speed of light in vacuum  \cite{cavities}, when they are actually confirming the Lorentz contraction effect for physical objects, such as rods and resonant cavities. This flaw explains the null results from LIGO and related gravitational detectors, despite the fact that other experimental techniques, such as the gas-mode laboratory Michelson interferometers and the one-way RF coaxial cable experiment by DeWitte \cite{DeWitte}, have repeatedly detected the space velocity fluctuations - the phenomena underlying gravitational waves. The design flaw can be overcome by using a gas  or other dielectric in the light paths, as first reported in  2002 \cite{MMCK}.  However LISA, which is essentially a space-based vacuum-mode Michelson interferometer, and so without rods forming the arms,  does not experience a Lorentz contraction of the ``arms".  Repeating the above analysis gives for the difference in travel times, assuming for simplicity equal length arms,
\begin{equation}\Delta t=t_{AC}-t_{\perp AC}=\frac{L\cos(2\theta)v^2}{c^3} \mbox{\   to  \ } O(v^2/c^2)\label{eqn:dt}\end{equation}
Then, because $L$ is  planned to be some   $5\times 10^6$ km,  LISA will be  ultra-sensitive, with excessive effective fringe shifts overwhelming the detection system.   

Fortunately early experiments had a gas present, so that $n\neq 1$, albeit very close to $1$.  As well some used air while others used helium, and only by taking account of the different refractive indices does the data become consistent, see \cite{MMCK}.  For dielectrics with $n$ not near 1, the Fresnel drag effect must be taken into account, see \cite{CBFresnel} for a derivation of this effect, and \cite{Book} for an early discussion of its role in interferometers.

The results from the laboratory gas-mode Michelson interferometers are only consistent with the spacecraft earth-flyby Doppler shift data if the Lorentz contraction involves the speed of the rods wrt to space, as in Lorentz Relativity,  of some 480km/s \cite{CahillNASA}, and not the speed of the rods wrt the observer, as in Einstein Special Relativity. So together these experiments distinguish the very different Lorentzian and Einsteinian accounts of relativistic effects.

\section{Conclusions}
Misunderstandings about the anisotropy of the speed of light have confounded physics for more than 100 years.     That happened  because Michelson, understandably, assumed the correctness of Newtonian physics in calibrating his interferometer. Later  Fitzgerald and Lorentz introduced the notion of a physical length contraction of rods when in motion through space, but did not re-calibrate the interferometer in order to understand the significance of the small but not null data.  Only in 2002 \cite{MMCK} was this oversight of the contraction effect first corrected in analsying data from the early gas-mode interferometers. In the case of \cite{CH} assumptions contrary to known experimental outcomes 
were made, leading to a spurious and false claim.  The existence of a dynamical space has lead to a major  development of a new physics \cite{Book,CahillMink,Review}.

\end{document}